\documentclass{webofc}

\usepackage[varg]{txfonts}   
\usepackage{hyperref}
\usepackage{url}
\hypersetup{colorlinks=true,citecolor=blue,urlcolor=blue,linkcolor=blue}
\begin{document}
\title{First $D^0+\overline{D}^0$ measurement in heavy-ion collisions at SPS energies with NA61/SHINE}
%
%

\author{\firstname{Anastasia} \lastname{Merzlaya}\inst{1}\fnsep\thanks{\email{anastasia.merzlaya@cern.ch}}  for the NA61\slash SHINE Collaboration
}

\institute{University of Oslo         }

\abstract{
The measurement of open charm meson production provides a tool for the investigation of the properties of the hot and dense matter created in nucleus-nucleus collisions at relativistic energies. In particular, charm mesons are of vivid interest in the context of the study of the nature of the phase-transition between confined hadronic matter and the quark-gluon plasma. Recently, the experimental setup of the NA61/SHINE experiment was upgraded with the high spatial resolution Vertex Detector which enables the reconstruction of secondary vertices from open charm meson decays.

In this presentation the first $D^0$ meson yields at the SPS energy regime will be shown. The analysis used the most central 20\% of  Xe+La collisions at 150A GeV/c from the data set collected in 2017. This allowed the estimation of the corrected yields (dN/dy) for $D^0+\overline{D}^0$ via its $\pi^{+/-} + K^{-/+}$ decay channel at mid-rapidity in the center-of-mass system. The results will be compared and discussed in the context of several model calculations including statistical and dynamical approaches

}
\maketitle
\section{Introduction}
\label{intro}
	The study of open charm production is a sensitive tool for detailed investigations of the properties of hot and dense matter formed in nucleus-nucleus collisions. 
	Since heavy quarks are produced in hard scattering processes in the early stage of the collision, by studying them one can get insight to the properties of the created medium.
	In particular, such measurements are of vivid interest at the SPS energies, which are close to the quark-gluon plasma creation threshold.
		
	Also such study gives a unique opportunity to test the validity of theoretical models based on perturbative Quantum Chromodynamics and Statistical model approaches for nucleus-nucleus collisions at SPS energies, which provide very different predictions for charm yields \cite{NA61proposal}.

However, up to now there were no open charm measurements at this energy regime.

\section{Open charm measurements at NA61\slash SHINE }
	\label{sec-1}
	The SPS Heavy Ion and Neutrino Experiment  (NA61\slash SHINE) \cite{NA61} at CERN was designed for studies of the properties of the onset of deconfinement and search for the critical point of strongly interacting matter by investigating p+p, p+A and A+A collisions at different beam momentum from 13$A$ to 158$A$ GeV/c for ions and upto 400 GeV/c for protons. The main tracking detectors are four Time Projection Chambers (TPC) which provide momentum, charge and particle identification (PID) for the tracks produced in the collision.
	
	The heavy-ion programme of the NA61\slash SHINE experiment at CERN SPS has been expanded to allow precise measurements of particles with short lifetime.  To meet the challenges of open charm measurements, NA61\slash SHINE was upgraded with the Small Acceptance Vertex Detector (SAVD).  
SAVD is composed of four detector planes located $5, 10, 15$ and $20$ cm downstream of the target. The planes are constructed from position-sensitive MIMOSA-26AHR pixel sensors \cite{mimosa}.  SAVD provides precise spatial resolution of vertex reconstruction. The obtained transverse and longitudinal primary vertex resolution is $\sigma_{x,y} \approx$ 1 $\mu$m and $\sigma_z$ = 15 $\mu$m, respectively. The SAVD acceptance for $D^0,~\overline{D^0}$  is  $-0.5<y<1.0,~ 0.2<p<2.0~GeV$, and the analysis presented here is performed in this kinematic range.
More details about SAVD, track and vertex reconstruction can be found in Refs.~\cite{Merzlaya:2771816,Aduszkiewicz:2023qac}.
	
The $D^0+\overline{D}^0$ reconstruction is done via $\pi^{+/-} + K^{-/+}$ decay channel.
The SAVD tracks matched to TPC tracks are used to search for the $D^0 + \overline{D}^0$ signal. 
In the analysis presented here, the PID information was not used.
Each track is paired with all the opposite-sign tracks from the same event and is assumed to be either a kaon or a pion, thus each pair contributes twice in the combinatorial invariant mass distribution.
In order to suppress the combinatorial background of the invariant mass distribution, the following cuts are applied:
\begin{itemize}
\item cut on the daughter track impact parameter, $d>36~\mu$m;
\item cut on the distance of closest approach between the daughter tracks, $DCA < 42~\mu$m;
\item cut on the longitudinal distance between the $D^0$ decay 
vertex candidate and the primary vertex scaled with the gamma factor, $V_z/\gamma > 0.15~\mu$m;
\item cut on the impact parameter of the back-extrapolated $D^0$ candidate momentum vector, $D~<~20~\mu$m;
\item cut on the $D^0$ candidate momentum, $13 < p < 38$ GeV/$c$.
\end{itemize}

\begin{figure}
\centering
\sidecaption
\includegraphics[width=7.5cm,clip]{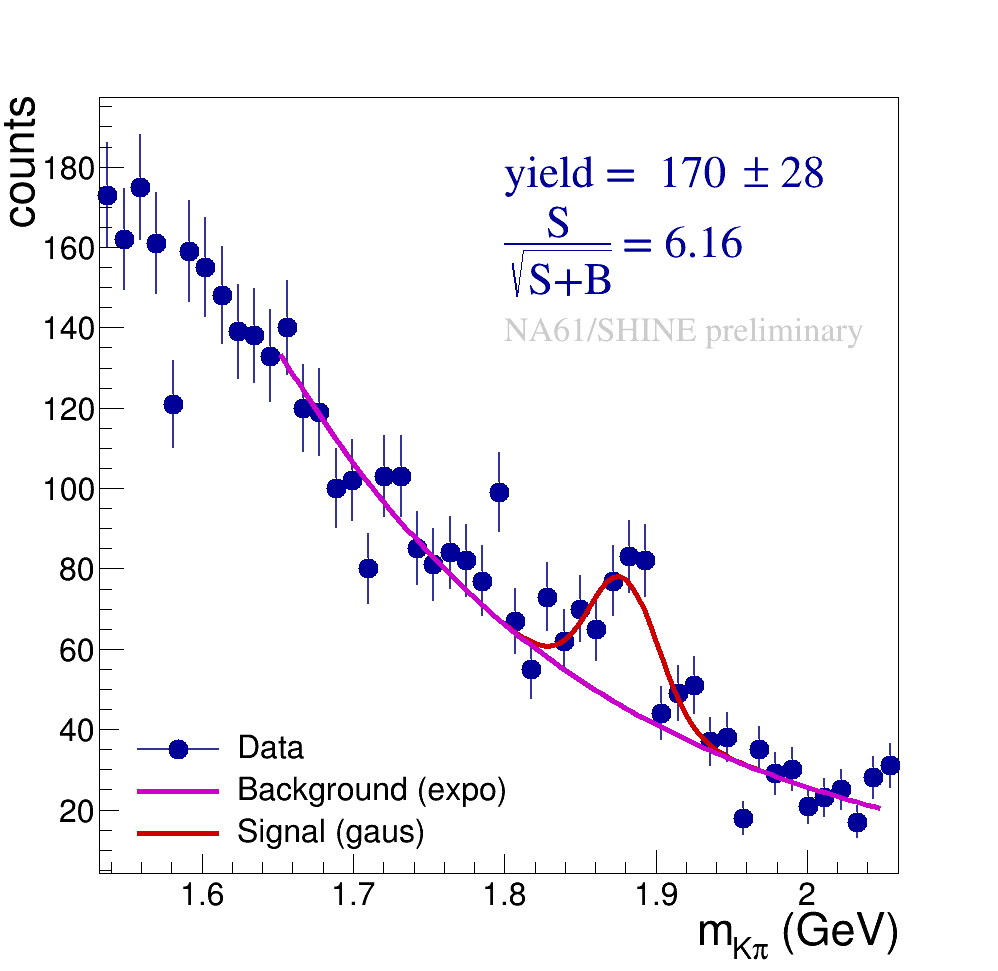}
\caption{Invariant mass distribution of unlike charge sign $\pi, K$ decay track candidates  for Xe+La collisions at 150$A$ GeV/$c$ taken in 2017.
The invariant mass distribution was fitted (red line) using an exponential function to describe the background and a Gaussian to describe the $D^0 + \overline{D}^0$ signal contribution.
The indicated errors are statistical only.
}
\label{fig:Figure_D0}
\end{figure}

Figure ~\ref{fig:Figure_D0} shows the invariant mass distribution of unlike charge daughter candidates with the applied cuts.
The peak corresponding to $D^0 + \overline{D}^0$ has a statistical significance, $S/\sqrt{S+B}$, of approximately 6$\sigma$. 

\section{Analysis results}
	\label{sec-2}

In order to obtain the corrections and calculate the absolute value of $D^0 + \overline{D}^0$ yield, the GEANT4 simulations were performed. 
The background in the Monte Carlo (MC) event was described using the EPOS model \cite{Pierog:2013ria},
while the signal phase space was parametrized using 3 models: AMPT \cite{Lin:2004en}, PHSD \cite{Cassing:2009vt} and PYTHIA/Angantyr \cite{Bierlich:2018xfw}, which predict quite different phase space distribution of open charm.

After applying corrections from MC, the corrected yield for each phase space assumption was calculated. The results are presented in the Table~\ref{tab:table_D0Results}. The visible yield $N(D^0+\overline{D}^0)_{visible}$ corresponds to the yield in the $-0.5<y<1.0,~ 0.2<p<2.0~GeV$ rapidity-transverse momentum range covering SAVD acceptance. In order to obtain the rapidity density, $dN(D^0+\overline{D}^0)/dy~_{-0.5<y<1.0}$, and the $4\pi$ integrated yield, $\langle D^0+\overline{D}^0\rangle$, the extrapolation factors for each of the models were determined. The AMPT model predicts that about 84\% of $D^0 + \overline{D}^0$ are produced in the SAVD acceptance, while PHSD and PYTHIA/Angantyr predict this fraction to be approximately 67\%. Thus, $\langle D^0+\overline{D}^0\rangle$ differ significantly between the model assumptions.

\begin{table}
    \centering
    \caption{Results for the visible yield, $dN/dy$ at mid-rapidity and yield in $4\pi$ of $D^0+\overline{D}^0$ obtained assuming the phase space distribution for AMPT, PHSD and PYTHIA/Angantyr models. The first indicated error corresponds to statistical uncertainty and the second - to systematically. Systematical uncertainty doesn't include the correction to model-dependent phase space.}
    \label{tab:table_D0Results}
    \begin{tabular}{c|c|c|c}
        \hline
        correction with: &  $N(D^0+\overline{D}^0)_{visible}$ & $\frac{dN(D^0+\overline{D}^0)}{dy}_{-0.5<y<1.0}$ & 
        $\langle D^0+\overline{D}^0\rangle$\\
        \hline
        AMPT  & 0.184 $\pm$ 0.032 & 
        0.129 $\pm$ 0.023 $\pm$ 0.035 & 
        0.218 $\pm$ 0.039 $\pm$ 0.060 \\
        PHSD & 0.204 $\pm$ 0.036 & 
        0.148 $\pm$ 0.026 $\pm$ 0.036  & 
        0.303 $\pm$ 0.054 $\pm$ 0.074 \\
        PYTHIA/Angantyr & 0.201 $\pm$ 0.035  & 
        0.147 $\pm$ 0.026 $\pm$ 0.037  &
        0.300 $\pm$ 0.052 $\pm$ 0.075 \\
    \end{tabular}
\end{table}

\begin{figure}[h]
\centering
\sidecaption
\includegraphics[width=12cm,clip]{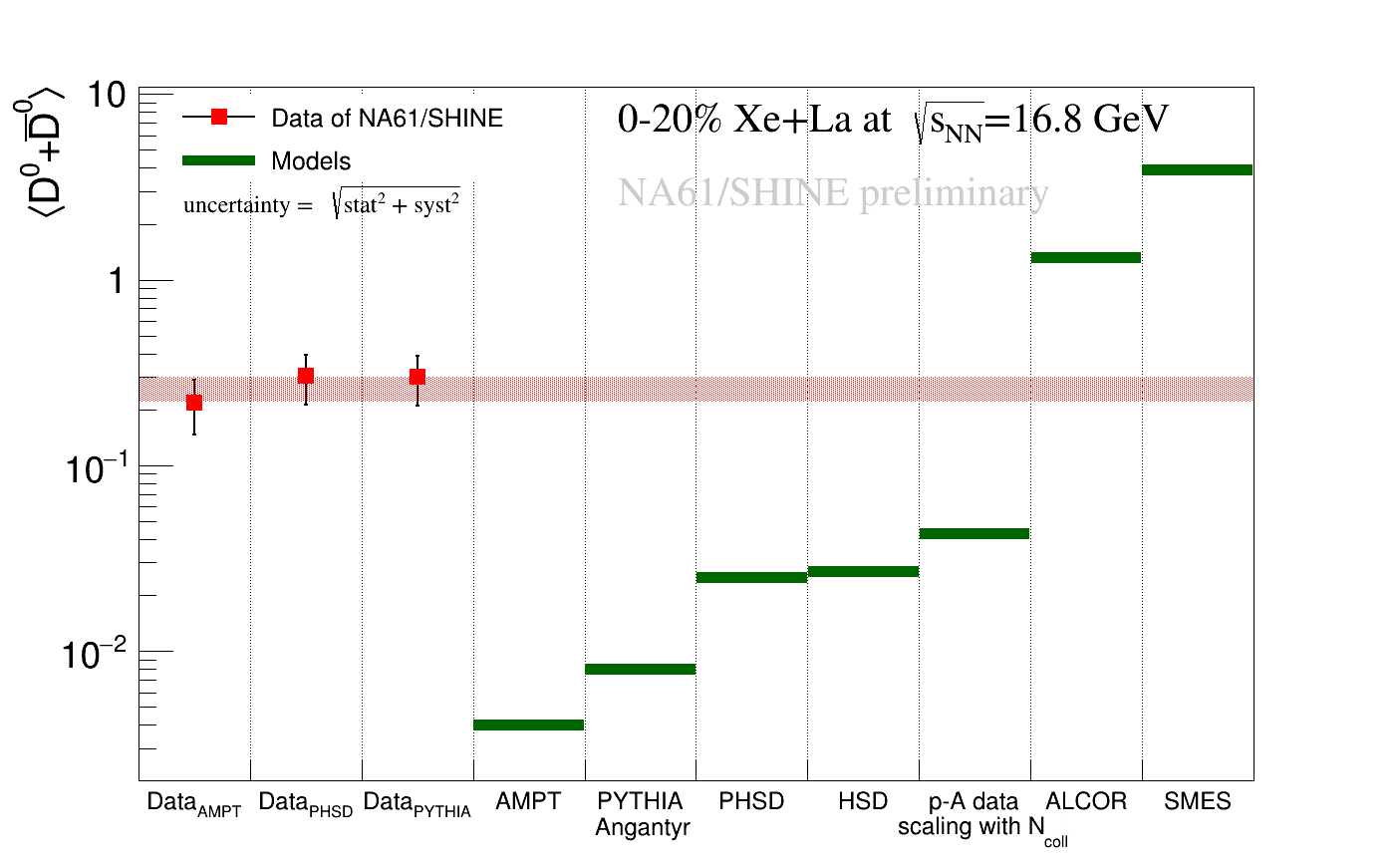}
\caption{Comparison of the obtained result to theoretical model predictions. The red band indicates the theoretical uncertainty of the result due to the unknown phase space distribution of $D^0,~\overline{D}^0$.}
\label{fig:Figure_models}
\end{figure}

Fig.\ref{fig:Figure_models} shows the comparison of the obtained $4\pi$ yield $\langle D^0+\overline{D}^0\rangle$ to the theoretical model predictions. The precision of the data is sufficient to discriminate between the current model predictions. While microscopic models (AMPT, PYTHIA/Angantyr, PHSD, HSD\cite{Linnyk:2008hp}) tend to significantly underestimate  $D^0+\overline{D}^0$ yield, ALCOR \cite{Levai:2000ne} and SMES \cite{Gazdzicki:1999mc} models are overestimating it. 

An estimation based on measurements of the $D^0+\overline{D}^0$ cross-section in different p+A collision systems and in a range of collision energies, $\sqrt{s_{NN}}$, spanning between 20 and 40 GeV \cite{Braun-Munzinger:1997pue}, which are scaled by the number of binary collisions corresponding to central Xe+La collisions, yields a value that is below the NA61\slash SHINE measurement with a statistical significance of 2-3$\sigma$.

\section{Summary and outlook}
The presented analysis shows the first direct measurement of open charm at the SPS energies in heavy-ion collisions with the precision of the result sufficient being able to disentangle between the most extreme theoretical predictions.

Looking forward,  NA61\slash SHINE was upgraded during CERN long-shutdown 2 to increase the data taking rate from 80Hz to 1kHz \cite{NA61proposal}. Within the upgrade the new Vertex Detector based on ALPIDE sensors developed was installed, as well as the TPC readout and DAQ were being upgraded.
In 2022-2023 NA61\slash SHINE collected on the lever of 180M events of Pb+Pb collisions at 150$A$~GeV/$c$. 
	These data should allow rapidity and transverse momentum differential measurements of $D^0$ and $\overline{D}^0$, as well as measurements of other charm hadrons.
	This study will provide a better insight into charm production mechanisms at energies close to the production threshold.
	
~\

\begin{acknowledgement}
		This work was supported by the Norwegian Financial Mechanism 2014--2021 (grant 2019\slash 34\slash H\slash ST2\slash 00585).
\end{acknowledgement}

%
%
%

\end{document}